\documentclass{amsart}
\usepackage{graphicx}
\usepackage{amsmath}
\usepackage{amsthm}
\usepackage{mathtools}
\usepackage{amsfonts}
\usepackage{amssymb}
\usepackage{epstopdf}
\usepackage{caption}
\usepackage{mathtools}

\usepackage{tikz}
\usetikzlibrary{matrix,arrows,decorations.pathmorphing}
\usepackage{pgf}
\usepackage{amsmath,amsthm, amscd}
\usepackage{amsfonts}
\usepackage{txfonts}
\usepackage{marvosym}
\usepackage{graphicx}
\usepackage{color}
\usepackage{times}
\usepackage{epsfig}
\usepackage{setspace}
\usepackage{mathrsfs}
\begin{document}
\title{Note on minimal number of skewed unit cells for periodic distance calculation}
\author{Senja Barthel}
\email{senja.barthel@epfl.ch}

\begin{abstract}
How many copies of a parallelepiped are needed to ensure that for every point in the parallelepiped a copy of each other point exists, such that the distance between them equals the distance of the pair of points when the opposite sites of the parallelepiped are identified? This question is answered in Euclidean space by constructing the smallest domain that fulfills the above condition. We also describe how to obtain all primitive cells of a lattice (i.e., closures of fundamental domains) that realise the smallest number of copies needed and give them explicitly in 2D and 3D.
\end{abstract}

\maketitle
\section{Introduction}
\begin{figure}[h]
\vskip 1cm
\includegraphics[width=1\linewidth]{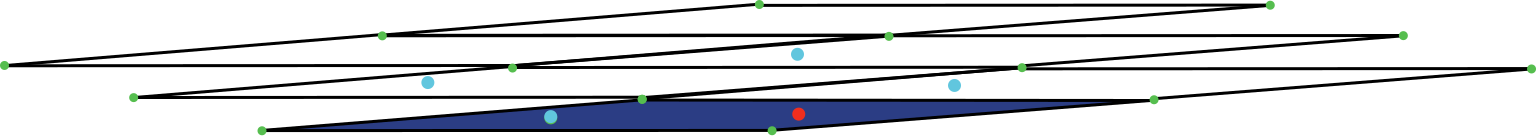}
\caption{$P$ in dark blue, $\Lambda$ in green. Skewed cells might require many copies until the smallest distance is obtained:  The copy of a blue point that is closest to the red point does not lie in a neighbouring cell of~$P$.}
\label{2}
\end{figure}

The question addressed in this paper appears when it is asked for distances between pairs of points in a parallelepiped $P$ whose opposite sides are identified. Such a periodic setting is natural in crystallography and can also be found in many other fields such as molecular biology, fluid mechanics, and astronomy. A basic operation that requires the computation of pairwise distances in a periodic setting is for example the computation of the Voronoi decomposition of a crystal where the centre points of the Voronoi cells are the atom positions. This is used for analyses as well as for defining the bond network of the molecule~\cite{Voro, Zeo, Topos, Blatov04}. It is common to add the neighbouring copies of the parallelepiped to compute the distance respecting periodic boundary conditions. That is, in 2D 9 copies and in 3D 27 copies are considered, in general $3^n$ copies are taken in $\mathbb{R}^n$ (Figure~\ref{1}). The distance between two points $p_{1}, p_{2} \in P$ is then computed as the minimal distance between $p_1$ and the periodic copies of $p_2$. Also most proved algorithms to compute Delaunay triangulations from point sets in 3D periodic space require 27 copies~\cite{proved3DDelauneytTrinagulation27cells} although it is possible to work on one parallelepiped only, if the parallelepiped is orthogonal~\cite{CgalVoro, 3DVoronoi}. However, for very tilted parallelepipeds, more than the neighbouring copies might be needed to be taken into account in order to compute all pairwise distances correctly with a procedure that is choosing the smallest distance between copies of points (Figure~\ref{2}). Luckily, for the reduced and conventional cells, which are usually given in crystallography, $3^n$ copies turn out to be sufficient. Pairwise distance calculations in periodic settings are also required when constructing the Vietoris-Rips or Alpha complex for a persistent homological analysis on a point cloud in a periodic setting. The parallelepipeds in this setting can be tilted, for example when they are constructed from crystallographic cells~\cite{poreshape}. 

\section{Number of copies}
Let $B:=(v_{1}, \dots, v_{n}) \in \mathbb{R}^{n^2}$ equipped with the $2$-norm be a matrix whose columns are $n$ linearly independent vectors and $P$ be the parallelepiped spanned by them (containing the boundary). Let $F$ be the space obtained from identifying opposite faces of the parallelepiped $P$, i.e., $F = \mathbb{R}^n / \Lambda$, where $\Lambda=(B \mathbb{Z}^n, +)$. This defines a covering with covering map $\pi: \mathbb{R}^n \rightarrow  F$. The distance $d(q_1, q_2)$ between the points $q_1, q_2 \in F$  is defined as the minimal distance in $\mathbb{R}^n$ between two points $ p_1 \in \pi^{-1}(q_1)$ and $ p_2 \in \pi^{-1}(q_2)$. 

We want to construct a connected compact minimal subset $D \subset \mathbb{R}^3$ with $P \subset D$ such that for any given point $p$ in $P$, there exists a point $\tilde{p} \in D$ with $d_{\mathbb{R}}(p, \tilde{p})=d(\pi(p), \pi(\tilde{p}))$, where $d_{\mathbb{R}}$ is the metric inherited from~$\mathbb{R}^n$.

\begin{figure}[h]
\includegraphics[width=0.4\linewidth]{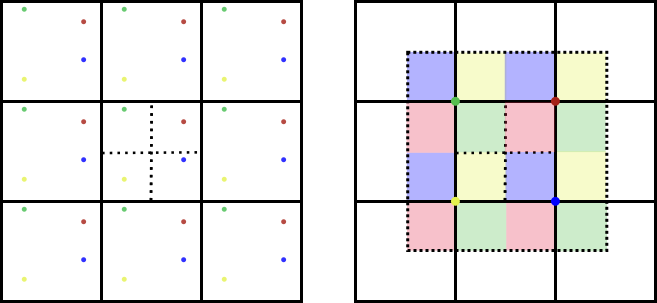}
\caption{Finding pairwise distances under periodic boundary
conditions. Orthogonal cells:  taking $3^n$~copies is sufficient (left); domains (green, yellow, purple, pink) of the middle cell $P$ that are closest to each corner point of $P$, and the copies of the domains attached to each corner point of $P$ (right).}
\label{1}
\end{figure}

\begin{figure}[h]
\vskip 0.5cm
\includegraphics[width=0.8\linewidth]{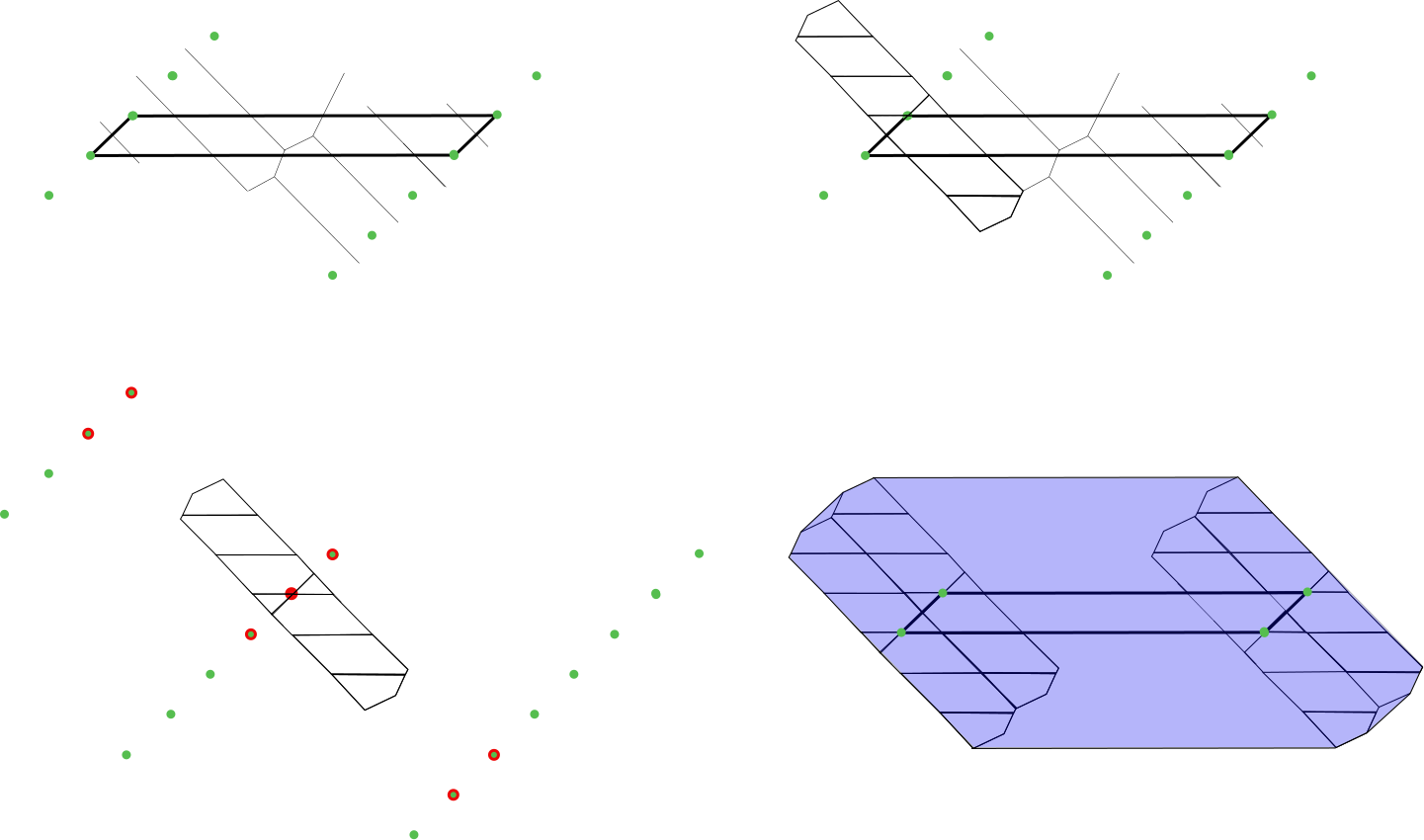}
\caption{Another example of a skewed cell. The lattice $\Lambda$ in green, Voronoi relevant points red encircled, and $D$ blue shaded.}
\label{3}
\end{figure}

For orthogonal $P \subset \mathbb{R}^n$, $D$ is clearly contained in $3^n$ copies of $P$, three in each direction of the linearly independent vectors that span $P$ (Figure~\ref{1}). However, more copies of $P$ are needed to cover $D$ in the general case of non-orthogonal vectors $v_{1}, \dots, v_{n}$. Such a case is shown in Figure~\ref{2} where $P$ is shaded dark blue: The light blue point closest to the red point is not contained in $P$ and the eight copies around it. We show how to obtain $D$ and illustrate the argument with examples (Figure~\ref{1},\ref{3}).
\vskip 8pt
\noindent
\textbf{Determining \textit{D}}\\
\textit{Let} V \textit{be a Voronoi cell with respect to the lattice} $\Lambda$. \\
D \textit{is the union of} P \textit{and copies of} V \textit{attached to each point of the boundary of} P.
\vskip 5pt
\noindent
\textbf{Proof:}\\
To construct $D$ as defined above, we have to add to each point $p \in P$ the closure of the set of points that are closer to $p$ than to any of its copies $\pi_{-1}(\pi(p)) \smallsetminus p$. This is by definition the Voronoi cell $V_p$ that contains $p$ of the set $\Lambda_p$, where the set $\Lambda_p$ is the translate of $\Lambda$ by the vector $p$. Denote the union of all Voronoi cells $V_q$ of points $q \in \partial P$ by  $U$. This is, $U$ is obtained by shifting $V$ along $\partial P$ since the Voronoi cells obtained from lattices that are related by a translation are translational equivalent. $U$ covers every Voronoi cell $V_p$, $p \in P$ by construction and it follows that $D \subset U$. Since all points of $\mathring{U}$ are closer to a point $p \in P$ than to any of its copies $\pi_{-1}(\pi(p)) \smallsetminus p$, it follows that $\mathring{U} \subset D$. Since both $U$ and $D$ are closed, we have shown that~$U=D$. \hskip 11.7cm $\square$
\vskip 8pt

To determine the minimal number of copies of $P$ that is needed such that $D$ is contained in their union, one only needs to project $D$ on the vectors that span $P$ and take the ceiling of the result divided by their lengths respectively. \\

\section{Choices of cells}
\begin{figure}[h]
\vskip 0.5cm
\includegraphics[width=0.4\linewidth]{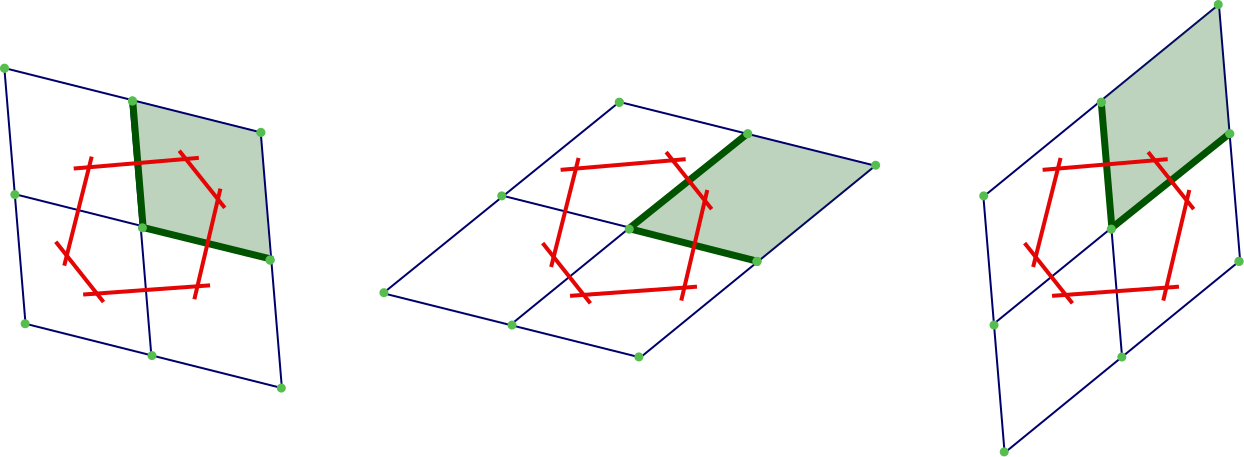}
\caption{The three choices for $P_s$ in 2D. Left to right: The domains given by $(v_{1}, v_{2})$, $(v_{1}, v_{1}+v_{2})$, and $(v_{1}+v_{2}, v_{2})$, where $v_{1}, v_{2}$ are the shortest vectors enclosing an angle of more than $90^\circ$ that form a basis of the lattice.}
\label{5}\vskip 1cm
\end{figure}

Note that for the distance computation it would suffice to consider the space that is obtained by removing one of each pair of opposite facets from the boundary of $D$. That allows to use fundamental domains of a lattice in the following instead of always writing `the closure of the fundamental domain' to address the corresponding parallelepiped. The pairwise distances in $\mathbb{R}^n / \Lambda$ are independent from the choice of the basis of the lattice $\Lambda$. Therefore, if we are interested in the pairwise distance between points in $F$ only, i.e., if $D$ is only used to calculate the distances in $F$ without the parallelepiped being of interest in its own, we can restrict our consideration to those fundamental domains $P_s$ of the lattice $\Lambda$, for which $2$ copies in each dimension around the origin cover the Voronoi cell $V$ of $\Lambda$. The above argument shows for that in this case $3^n$ copies will suffice to compute all pairwise distances under periodic boundary conditions correctly by taking the minimal distance between periodic copies of points. Denote the union of these $3^n$ copies of a fundamental domain $P_s$ by $C$. The $P_s$ are exactly those fundamental domains whose corresponding domains $C$ contain all Voronoi relevant points of $V_0$ in their boundary. In Euclidean space it is known which points of a lattice are Voronoi relevant~\cite{Minkowski}: In the orthogonal case there are $2n$ Voronoi relevant points, namely $\pm v_{1}, \dots, \pm v_{n}$, where $v_{1}, \dots, v_{n}$ are the $n$ shortest lattice vectors in the $n$ linearly independent directions. For non-orthogonal cases, there are $2(2^n -1)$ Voronoi relevant vectors. The Voronoi relevant vectors in 2D are $ \pm v_{1}$, $\pm v_{2}$, and $\pm (v_{1} + v_{2})$ if the angle between $v_{1}$ and $v_{2}$ is larger than $90^{\circ}$, respectively $ \pm v_{1}$, $\pm v_{2}$, and $\pm (v_{1} - v_{2})$ if the angle between $v_{1}$ and $v_{2}$ is less than $90^{\circ}$, where $v_{1}$ and $v_{2}$ are the shortest linearly independent lattice vectors. To determine all $P_s$ in 2D it is sufficient to consider $v_{1}, v_{2}$ such that their angle is larger than or equal to $90^{\circ}$: There are three choices of fundamental domains $P_s^1$, $P_s^2$, and $P_s^3$, spanned by $(v_{1}, v_{2})$, $(v_{1}, v_{1}+v_{2})$, and $(v_{1}+v_{2},v_{2})$ respectively (Figure~\ref{5}). All other domains obtained from different choices of Voronoi relevant points that form a basis of $\Lambda$ can be translated to one of $P_s^1$, $P_s^2$, or $P_s^3$. For example, $v_{1}, v_{2}$ and $v_{1}, -v_{2}$ span domains that are related by a translation.\\
The argument is similar in higher dimensions. Since 3D is relevant for crystallographic applications and we want to know for which choices of crystallographic unit cells 27 copies are sufficient for pairwise distances calculations, it is spelled out in the following: Again, let $v_{1}, v_{2}, v_{3}$ be the shortest linearly independent vectors generating the lattice with pairwise enclosed angles larger than or equal to $90^{\circ}$. The Voronoi relevant vectors are $\pm v_{1}$, $\pm v_{2}$, $\pm (v_{1}+v_{2})$, $\pm (v_{1}+v_{3})$, $\pm (v_{2}+v_{3})$, and $\pm (v_{1}+v_{2}+v_{3})$. The 19 possible choices of domains $P_s$ are the following, given by their three spanning vectors:
\vskip 5pt
 $\begin{array}{c c c c }
 (v_{1}, v_{2}, v_{3}) & (v_{1}, v_{2}, v_{1}+v_{3}) & (v_{1}, v_{2}, v_{2}+v_{3}) & (v_{1}, v_{2}, v_{1}+v_{2}+v_{3})\\
 (v_{1}, v_{1}+v_{2}, v_{3})&(v_{1}, v_{1}+v_{2}, v_{1}+v_{3})&(v_{1}, v_{1}+v_{2}, v_{1}+v_{2}+v_{3})&(v_{1}, v_{2}+v_{3}, v_{3})\\
 (v_{1}, v_{1}+v_{2}+v_{3},v_{3}) & (v_{1}, v_{1}+v_{2}+v_{3},v_{1}+v_{3}) & (v_{1}+v_{2}, v_{2}, v_{3}) & (v_{1}+v_{2}, v_{2}, v_{2}+v_{3}) \\
 (v_{1}+v_{2}, v_{2}, v_{1}+v_{2}+v_{3})&(v_{1}+v_{3}, v_{2}, v_{3})&(v_{1}+v_{3}, v_{1}+v_{2}+v_{3}, v_{3})&(v_{1}+v_{3}, v_{2}+v_{3}, v_{3})\\
 (v_{1}+v_{2}+v_{3}, v_{2}, v_{3}) & (v_{1}+v_{2}+v_{3}, v_{2}, v_{2}+v_{3}) & (v_{1}+v_{2}+v_{3}, v_{2}+v_{3}, v_{3})
 \end{array}$ \enlargethispage{2\baselineskip}
\vskip 5pt
In particular, if a parallelepiped is spanned by the shortest $n$ linearly independent lattice vectors $(v_{1}, v_{2}, v_{3})$, $2^n$ copies around the origin cover the Voronoi cell of the lattice and therefore $3^n$ copies are sufficient to determine the distances in $F$ correctly by taking the pairwise distances. This includes the reduced cells from crystallography as well as the conventional cells (which might contain several copies of primitive cells but are themselves reduced cells with respect to the sublattice that is generated by the cell vectors of the conventional cells). In most crystallographic settings it is therefore sufficient to take 27 copies (or 9 copies in 2D) to compute pairwise distances in a crystal correctly. Although the problem of finding a shortest basis is NP hard for general metrics~\cite{NPhard}, there is a single exponential time algorithm for the Euclidean space~\cite{shortest}. In non-Euclidean spaces, the number of Voronoi relevant vectors can be much larger~\cite{l3norm}.
\newpage

\begin{figure}[h]
\vskip 0.5cm
\includegraphics[width=1\linewidth]{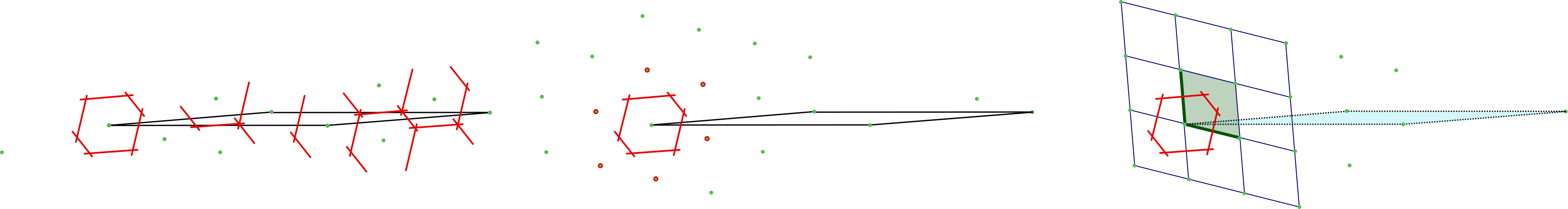}
\caption{The skewed parallelepiped from from Fig.~\ref{2} in shaded light blue. Red lines: enclosing Voronoi cells of the lattice. Voronoi relevant points of $V$ are encircled in red. Right: Shortest basis of the lattice in green (spanning a reduced cell).}
\label{4}
\end{figure}


\begin{thebibliography}{0}
\bibitem{Voro}C.H.~Rycroft, Voro++: A three-dimensional Voronoi cell library in C++. {\it Chaos,} {\bf 19}, 041111 (2009).

\bibitem{Zeo}T.F.~Willems, C.H.~Rycroft, M.~Kazi, J.C.~Meza, M.~Haranczyk, Algorithms and tools for high-throughput geometry-based analysis of crystalline porous materials. {\it Microporous and Mesoporous Materials,} {\bf 149}, 134--141 (2012) doi:10.1016/j.micromeso.2011.08.020. 

\bibitem{Topos}V.A.~Blatov, A.P.~Shevchenko, D.M.~Proserpio, Applied topological analysis of crystal structures with the program package ToposPro. {\it Cryst. Growth Des.,} {\bf 14}, 3576–-3586 (2014).

\bibitem{Blatov04}V.A.~Blatov, Voronoi–dirichlet polyhedra in crystal chemistry: theory and applications. {\it Crystallography Reviews,} {\bf 10:4}, 249--318 (2004) doi: 10.1080/08893110412331323170.

\bibitem{proved3DDelauneytTrinagulation27cells}N.~Dolbilin, D.~Huson, Periodic Delone Tilings. {\it Periodica Mathematica Hungarica,} {\bf 34}, 57--64 (1997) 10.1023/A:1004272423695.

\bibitem{CgalVoro}M.~Caroli, A.~Pell\'{e}, M.~Rouxel-Labb\'{e}, M.~Teillaud, 3D Periodic Triangulations. {\it In CGAL User and Reference Manual. CGAL Editorial Board,} {\bf 4.12 edition} (2018).

\bibitem{3DVoronoi}M.~Caroli, M.~Teillaud, Computing 3D Periodic Triangulations. {\it In: Fiat A., Sanders P. (eds) Algorithms - ESA 2009. ESA 2009. Lecture Notes in Computer Science,} {\bf 5757} Springer, Berlin, Heidelberg (2009).

\bibitem{poreshape}Y. Lee, S.D. Barthel, P. D\l{}otko, S.M. Moosavi, K. Hess, B. Smit, Quantifying similarity of pore-geometry in nanoporous materials. {\it Nature Communications,} {\bf 8} (2017)
doi: 10.1038/ncomms15396.

\bibitem{Minkowski}H. Minkowski, Allgemeine Lehrs\"{a}tze \"{u}ber die konvexen
Polyeder. {\it Nachrichten der K. Gesellschaft der Wissenschaften zu G\"{o}ttingen. Mathematisch-physikalische Klasse,} 198--219 (1897). 

\bibitem{NPhard}M. Ajtai. Generating hard instances of lattice problems. {\it Proceedings of the twenty-eighth annual ACM symposium on Theory of computing. Philadelphia, Pennsylvania, United States: ACM,} 99--108 (1996) doi: 10.1145/237814.237838.

\bibitem{shortest} D. Micciancio, P. Voulgaris, A  deterministic  single  exponential time  algorithm  for  most  lattice  problems  based  on  Voronoi  cell  computations. {\it  SIAM  Journal  on  Computing,} {\bf 42}, 1364--1391 (2013) doi: 10.1137/100811970.

\bibitem{l3norm} J. Bl\"{o}mer, K. Kohn, Voronoi Cells of Lattices with Respect to Arbitrary Norms. {\it  SIAM Journal on Applied Algebra and Geometry,} {\bf 2}, 314--338 (2018) doi:10.1137/17M1132045.

\vskip 1cm
\end{thebibliography}
\end{document}